\documentclass[twocolumn,aps,prl]{revtex4-1}

\usepackage{amsmath,amssymb,amsthm}
\usepackage{mathtools}
\usepackage{xspace}
\usepackage[usenames,dvipsnames,svgnames,table]{xcolor}
\usepackage[colorlinks,hyperindex,breaklinks]{hyperref}
\usepackage{braket}
\usepackage{hyphenat}
\usepackage{bbold}
\usepackage{enumerate}
\usepackage{tabularx}
\usepackage[verbose]{placeins}
\usepackage[hang,flushmargin]{footmisc}
\usepackage{comment}

\usepackage[ngerman,british]{babel}

\usepackage{color}
\def\tr{\textcolor{red}}
\def\tr{}

\newcounter{todoc}
\newcommand{\todoxxx}[1]{\addtocounter{todoc}{1}%
{\bfseries\color{YellowOrange}{\small\sf$[$(\arabic{todoc}) XXX #1$]$}}%
}
\renewcommand{\todoxxx}[1]{}

\DeclareMathOperator{\Tr}{Tr}

\renewcommand{\arraystretch}{1.6}
\newcommand{\dd}{\mathrm d}

\newcommand{\matr}[1]{\mathbf{\underline{#1}}}
\newcommand{\matrg}[1]{\underline{\boldsymbol{#1}}}

\newcommand{\ee}{\text e}
\newcommand{\ii}{\text i}

\renewcommand{\vec}[1]{\mathbf{#1}}
\newcommand{\spacegroup}[1]{$#1$}

\newcommand{\tableheading}[1]{\textbf{#1}}
\newcommand{\name}[1]{#1}

\newcommand{\matrixtablearraystretch}{0.9}
\newcommand{\matrixtablearraycolsep}{0.3pt}

\begin{document}

\title{Spin-orbit induced longitudinal spin-polarized currents in non-magnetic solids}

\author{S. Wimmer}
\email{sebastian.wimmer@cup.uni-muenchen.de}
\author{M. Seemann}
\author{K. Chadova}
\author{D. K\"odderitzsch}
\email{diemo.koedderitzsch@cup.uni-muenchen.de}
\author{H. Ebert}

\affiliation{Department Chemie/Phys.\,Chemie, 
             Ludwig-Maximilians-Universit\"at
             M\"unchen, Germany}

\date{\today}

\begin{abstract}
  For certain non-magnetic solids with low symmetry the occurrence of
  spin-polarized longitudinal currents is predicted. These arise due to
  an interplay of spin-orbit interaction and the particular crystal
  symmetry.
  This result is derived using a group-theoretical scheme that allows
  investigating the symmetry properties of any linear response tensor
  relevant to the field of spintronics. For the spin conductivity tensor
  it is shown that only the magnetic Laue group has to be considered in
  this context.
  Within the introduced general scheme also the spin Hall- and additional related
  transverse effects emerge without making reference to the two-current
  model.
  Numerical studies confirm these findings and demonstrate for
  (Au$_{1-x}$Pt$_{\rm x}$)$_4$Sc that the longitudinal spin conductivity
  may be in the same order of magnitude as the conventional transverse
  one. The presented formalism only relies on the magnetic space group and
  therefore is universally applicable to any type of magnetic order.
\end{abstract}


\maketitle

The discovery of the spin Hall effect\,\cite{DP71,Hir99,SCN+04} (SHE) with its
particular feature of converting a longitudinal charge current into a
transverse spin current has sparked numerous studies that finally led to a deep
understanding of many effects that are spin-orbit induced. Among them are the
enigmatic anomalous Hall effect (AHE) that shares the same origin as the SHE
and many new phenomena emerging from a coupling of spin-, charge- and orbital
degrees of freedom in electric fields as well as temperature gradients.
Examples for these are the Edelstein effect (EE\,\cite{AL89a,Ede90}), as well
as the spincaloritronic pendants to the SHE and AHE, namely the spin- and
anomalous Nernst effects (SNE\,\cite{TGFM12,WKCE13}, ANE
\,\cite{WFBM13,WKE14}), respectively. 
Many models have been formulated that aim to capture particular contributions
to theses effects. For instance, the concept of the semiclassical Berry phase
that can be determined on the basis of the band structure of perfect crystalline
systems is connected to so called intrinsic
contributions\,\cite{FNT+03,YF05,NSO+10}. Extrinsic contributions arising from
scattering at impurities in non-perfect systems can, for example, be obtained
from diagrammatic methods\,\cite{Sin08} or Boltzmann transport
theory\,\cite{GFZM10}. 

The aforementioned transport phenomena and their different contributions
being linear in the driving fields should, in principle, be described using the
fundamental Kubo formula for the corresponding response function\,\cite{Kle66},
\begin{equation}
  \tau_{ij}(\omega) = \int_0^\infty \!\!\!\! \dd t 
\ee^{-\ii\omega t} \int_0^\beta\!\!\!\! \dd\lambda \Tr\left[\rho \hat A_j \hat
B_i(t+\ii\hbar\lambda) \right]
  \, .
  \label{eq:Kubo-formula}
\end{equation}
The effects then emerge from the characteristics of the underlying
Hamiltonian, the pair of chosen operators for perturbation ($\hat A_j$) and
observable ($\hat B_i$) and the symmetry of the system.
Due to the intractability of the problem to exactly solve the Kubo formula
for a realistic system in practice one has to recourse to approximations and/or
models. 
However, irrespective of this problem one can still analyze the
transformation properties of response tensors $\matrg\tau$ determined by the
Kubo formula to make statements which effects are in principle allowed, i.e.\
which non-vanishing tensor elements may occur given a particular transformation
property of the operators appearing in Eq.~(\ref{eq:Kubo-formula}).
This route has been followed by Kleiner\,\cite{Kle66,Kle67}, who
demonstrated that the  occurrence of the AHE is predicted by such a space-time
symmetry analysis. 
Furthermore, considering in addition heat currents he derived general Onsager 
reciprocity relations.

Here, by extending this approach and applying it in the context of spin current 
operators\,\footnote{More precisely, one should use the term spin-polarization 
current density operator, but we keep the abbreviated term also in the 
relativistic context.} we demonstrate that in certain non-magnetic low symmetry 
systems an electric field can induce a \emph{longitudinal} spin polarized 
current\,\cite{SEK14} that has hitherto evaded perception, and complements the 
transverse spin Hall effect. Furthermore two additional transverse effects 
are found which differ from the SHE by the direction of polarization. The results 
of the group theoretical analysis are independently verified for an alloy bulk system 
performing relativistic first-principles Kubo-type transport calculations.
The presented formalism is furthermore very general, because (i) it allows
to identify other response phenomena as non-zero elements in respective
response tensors, as e.g. the AHE, (ii) it applies to both, magnetic and 
non-magnetic systems, and (iii) it is free of the notion of a two-current model 
often used as an approximation in discussing spintronic phenomena; instead it 
is based on the concept of spin (polarization) current densities.

The material-specific features of any  transport property may be discussed on
the basis of the corresponding response function tensor $\matrg \tau$.
Concerning this, the shape of the tensor $\matrg \tau$, i.e.\ the occurrence
and degeneracy of non-zero elements,  reflecting the symmetry of the
investigated solid, is obviously of central importance. To find, in particular,
the shape of the spin conductivity tensor, Kleiner's scheme\,\cite{Kle66} to
deal with the symmetry properties of ordinary transport tensors has been
extended  to the case when the response observable is represented by an arbitrary
operator product of the form  $(\hat B_i \hat C_j)$ while an operator $\hat A_k$ 
represents the perturbation and the operators  $\hat A_k$,  $\hat B_j$, and 
$\hat C_i$ are seen as  the Cartesian components of vector operators. Within 
Kubo's linear response formalism the corresponding frequency ($\omega$) 
dependent response function    $\matrg \tau$ is then given by
\begin{widetext}
\vspace*{-0.25cm}
\begin{align}\begin{split}
  \tau_{(\hat B_i \hat C_j) \hat A_k}(\omega, \vec H) = \int_0^\infty \dd t \:
\ee^{-\ii\omega t} \int_0^\beta \dd\lambda \Tr\left(\rho(\vec H) \hat A_k \hat
B_i(t+\ii\hbar\lambda; \vec H) \hat C_j(t+\ii\hbar\lambda; \vec H)\right)
  ,
  \label{eq:kleiner_transport_coefficient_three}
\end{split}\end{align}
\vspace*{-0.25cm}
\end{widetext}
where as usual\,\cite{Kle66} $\rho$ stands for the density operator, $\beta
= 1/k_BT$ with $k_B$ the Boltzmann constant, $T$ is the temperature and $\vec H$
is a magnetic field that might be present.

The shape of  $\matrg \tau$ can be found by considering the impact of a
symmetry operation of the  space group of the solid on
Eq.~\eqref{eq:kleiner_transport_coefficient_three}, as this will lead to
equations connecting elements of  $\matrg \tau$. Collecting the restrictions
imposed by all symmetry operations the shape of  $\matrg \tau$ is obtained.  In
this context it is important to note that the relevant space group of the
considered system may contain not only unitary pure spatial~($u$) but also
anti-unitary symmetry operations ($a$) that involve time reversal.

The transformation properties of the operators $ X =  A_i$, $ B_i$ or $ C_i$
in Eq.~\eqref{eq:kleiner_transport_coefficient_three} under symmetry operations
can be expressed in terms of the corresponding Wigner D-matrices\,\cite{Kle66}
$\matr D^{(\hat X)}(u)$ and $\matr D^{(\hat X)}(a)$ belonging  to the operator
$\hat X$ and the operation $u$ or $a$, respectively.  Starting from
Eq.~\eqref{eq:kleiner_transport_coefficient_three} and making use of these
transformation relations one gets the transformation behavior of $\matrg \tau$
under a unitary ($u$) or anti-unitary ($a$) symmetry operation, 
respectively:
%
\begin{eqnarray}
  \tau_{(\hat B_i \hat C_j) \hat A_k}(\omega, \vec H)
& =&
\sum_{lmn}
  \tau_{(\hat B_m \hat C_n) \hat A_l}(\omega, \vec H_u)
\nonumber  \\
&& \hspace{0cm}
\,  D^{(\hat A)}(u)_{lk}
\,  D^{(\hat B)}(u)_{mi}
\,  D^{(\hat C)}(u)_{nj}
  \label{eq:tau_trans_u}
\\
%
%
  \tau_{(\hat B_i \hat C_j) \hat A_k}(\omega, \vec H) & = &
 \sum_{lmn}
\tau_{\hat A_l^\dagger
      (\hat B_m^\dagger
      \hat C_n^\dagger)
       }(\omega, \vec H_a)
\nonumber  \\ && \hspace{-0.2cm}
\, D^{(\hat A)}(a)_{lk}^*
\, D^{(\hat B)}(a)_{mi}^*
\, D^{(\hat C)}(a)_{nj}^*
 \;  .
  \label{eq:tau_trans_a}
\end{eqnarray}
%
It should be   noted that in general the tensors 
$\tau_{(\hat B_i \hat C_j) \hat A_k}$ and 
$\tau_{\hat A_k^\dagger (\hat B_i^\dagger \hat C_j^\dagger)}$
are different objects representing different response functions which are only
interrelated by Eq.~\eqref{eq:tau_trans_a}.  It nevertheless imposes
restrictions on the shape of $\tau_{(\hat B_i \hat C_j) \hat A_k}$ giving rise
to (generalized) Onsager relations.

Assuming $\hat C_i = 1$ and $\hat B_i = \hat A_i = \hat j_i $ with $\hat j_i$ 
the current density operator $\matrg \tau$ corresponds to the ordinary
electrical conductivity tensor $\matrg \sigma$.  Using the behavior of  $\hat
j_i   $ under symmetry operations\,\cite{Kle66}, it turns out that only the
magnetic Laue group of the system has to be considered, that is generated by
adding the (space) inversion operation $I$ to the crystallographic magnetic
point group\,\footnote{In contrast to Kleiner we adopt here the definition of a
Laue group that is in general use nowadays.}.  The resulting shape of the
conductivity tensor  $\matrg \sigma$ is given in Table \ref{tab-TENSOR} for
four different magnetic Laue groups.
\begin{center}
\begin{table*}[htb]
  \renewcommand{\arraystretch}{\matrixtablearraystretch}
  \setlength\arraycolsep{\matrixtablearraycolsep}
  \centering
\begin{tabular}{>{\centering}m{4.0cm}||c|ccc}
\hline \hline\\
\tableheading{magnetic \name{Laue} group} & \tableheading{$\matrg \sigma$} & \tableheading{$\matrg \sigma^x$} & \tableheading{$\matrg \sigma^y$} & \tableheading{$\matrg \sigma^z$} \\[2mm]
\hline \hline
& & & \\
 \spacegroup{m\bar 3m1'}\\ \tr{(fcc-Pt)} & \begin{math} \begin{pmatrix} \sigma_{\rm xx} && 0 && 0\\0 && \sigma_{\rm xx} && 0\\0 && 0 && \sigma_{\rm xx} \end{pmatrix}\end{math} & \begin{math} \begin{pmatrix} 0 && 0 && 0\\0 && 0 && \sigma^{\rm x}_{\rm yz}\\0 && -\sigma^{\rm x}_{\rm yz} && 0 \end{pmatrix}\end{math} & \begin{math} \begin{pmatrix} 0 && 0 && -\sigma^{\rm x}_{\rm yz}\\0 && 0 && 0\\\sigma^{\rm x}_{\rm yz} && 0 && 0 \end{pmatrix}\end{math} & \begin{math} \begin{pmatrix} 0 && \sigma^{\rm x}_{\rm yz} && 0\\-\sigma^{\rm x}_{\rm yz} && 0 && 0\\0 && 0 && 0 \end{pmatrix}\end{math}\\[4mm]
& & & \\
 \spacegroup{4/mm'm'}\\ \tr{(fcc-Fe$_x$Ni$_{1-x}$)}
 & \begin{math} \begin{pmatrix} \sigma_{\rm xx} && \sigma_{\rm xy} && 0\\-\sigma_{\rm xy} && \sigma_{\rm xx} && 0\\0 && 0 && \sigma_{\rm zz} \end{pmatrix}\end{math} & \begin{math} \begin{pmatrix} 0 && 0 && \sigma^{\rm x}_{\rm xz}\\0 && 0 && \sigma^{\rm x}_{\rm yz}\\\sigma^{\rm x}_{\rm zx} && \sigma^{\rm x}_{\rm zy} && 0 \end{pmatrix}\end{math} & \begin{math} \begin{pmatrix} 0 && 0 && -\sigma^{\rm x}_{\rm yz}\\0 && 0 && \sigma^{\rm x}_{\rm xz}\\-\sigma^{\rm x}_{\rm zy} && \sigma^{\rm x}_{\rm zx} && 0 \end{pmatrix}\end{math} & \begin{math} \begin{pmatrix} \sigma^{\rm z}_{\rm xx} && \sigma^{\rm z}_{\rm xy} && 0\\-\sigma^{\rm z}_{\rm xy} && \sigma^{\rm z}_{\rm xx} && 0\\0 && 0 && \sigma^{\rm z}_{\rm zz} \end{pmatrix}\end{math}\\[4mm]
& & & \\
  \spacegroup{4/m1'}\\ \tr{(Au$_4$Sc)}
 & \begin{math} \begin{pmatrix} \sigma_{\rm xx} && 0 && 0\\0 && \sigma_{\rm xx} && 0\\0 && 0 && \sigma_{\rm zz} \end{pmatrix}\end{math} & \begin{math} \begin{pmatrix} 0 && 0 && \sigma^{\rm x}_{\rm xz}\\0 && 0 && \sigma^{\rm x}_{\rm yz}\\\sigma^{\rm x}_{\rm zx} && \sigma^{\rm x}_{\rm zy} && 0 \end{pmatrix}\end{math} & \begin{math} \begin{pmatrix} 0 && 0 && -\sigma^{\rm x}_{\rm yz}\\0 && 0 && \sigma^{\rm x}_{\rm xz}\\-\sigma^{\rm x}_{\rm zy} && \sigma^{\rm x}_{\rm zx} && 0 \end{pmatrix}\end{math} & \begin{math} \begin{pmatrix} \sigma^{\rm z}_{\rm xx} && \sigma^{\rm z}_{\rm xy} && 0\\-\sigma^{\rm z}_{\rm xy} && \sigma^{\rm z}_{\rm xx} && 0\\0 && 0 && \sigma^{\rm z}_{\rm zz} \end{pmatrix}\end{math}\\[4mm]
& & & \\
 \spacegroup{2/m1'}\\ \tr{(Pt$_3$Ge)} & \begin{math} \begin{pmatrix} \sigma_{\rm xx} && \sigma_{\rm xy} && 0\\\sigma_{\rm xy} && \sigma_{\rm yy} && 0\\0 && 0 && \sigma_{\rm zz} \end{pmatrix}\end{math} & \begin{math} \begin{pmatrix} 0 && 0 && \sigma^{\rm x}_{\rm xz}\\0 && 0 && \sigma^{\rm x}_{\rm yz}\\\sigma^{\rm x}_{\rm zx} && \sigma^{\rm x}_{\rm zy} && 0 \end{pmatrix}\end{math} & \begin{math} \begin{pmatrix} 0 && 0 && \sigma^{\rm y}_{\rm xz}\\0 && 0 && \sigma^{\rm y}_{\rm yz}\\\sigma^{\rm y}_{\rm zx} && \sigma^{\rm y}_{\rm zy} && 0 \end{pmatrix}\end{math} & \begin{math} \begin{pmatrix} \sigma^{\rm z}_{\rm xx} && \sigma^{\rm z}_{\rm xy} && 0\\\sigma^{\rm z}_{\rm yx} && \sigma^{\rm z}_{\rm yy} && 0\\0 && 0 && \sigma^{\rm z}_{\rm zz} \end{pmatrix}\end{math}\\[4mm]
\end{tabular}
  \caption{Electrical ($\matrg \sigma$) and spin ($\matrg \sigma^k$) conductivity tensor forms for the
magnetic Laue groups discussed in the text\,\cite{SEK14}. \tr{Below each group symbol an example for a material is given in parentheses.}}
  \label{tab-TENSOR}
\end{table*}
\end{center}
\vspace*{-0.85cm}
When considering the spin conductivity tensor its elements $\sigma_{ij}^k$ give
the  current density along direction $i$ for the spin polarization with respect
to the $k$-axis induced by an electrical field along the $j$-axis.  In this
case the perturbing electric field is still represented by 
$ \hat A_i  = \hat j_i  $  while the induced spin current density is represented by the
corresponding operator ${\hat J_i^k}=(\hat B_i  \hat C_k)$.  As the explicit
definition of ${\hat J_i^k}$ is not relevant for the following, but only its
symmetry properties, the frequently used non-relativistic definition $\hat
J_i^k=\frac{1}{2}\{\hat v_i,\sigma_k\}$ may be used that consists in a
combination of the Pauli spin matrix $\sigma_k$ and the conventional velocity
operator $\hat v_i$\,\cite{SZXN06}.  Alternatively, one may use the
relativistic definition of the spin current operator $\hat J_i^k = \hat {\cal
T}_k \hat j_i $ as suggested by Vernes \emph{et al}.\ \,\cite{VGW07} that
involves the spatial part $\hat {\cal T}_k  $ of the spin polarization
operator\,\cite{BW48}.

Expressing the transformation behavior of ${\hat J_i^k}$  in terms of the
Wigner matrices allows deducing the shape of the corresponding spin
conductivity tensor on the basis of Eqs.\ (\ref{eq:tau_trans_u}) and
(\ref{eq:tau_trans_a}).  As for the electrical conductivity it turns out again
that one has to consider only the magnetic Laue group, i.e.\  there are only 37
different cases.  Table \ref{tab-TENSOR} gives for the four cases considered
here the shape of the various sub-tensors  $\matrg \sigma^k$, where $k$
specifies the component of the spin polarization.

Considering a non-magnetic metal with fcc- or bcc-structure (\spacegroup{m\bar
3m1'}) Kleiner's scheme naturally leads to an isotropic electrical
conductivity tensor  $\matrg \sigma$.  The extension to deal with the spin
conductivity tensor sketched above gives in this case only few non-vanishing
elements that are associated with the SHE and are symmetry related according
to: $\sigma^{\rm x}_{\rm yz} = \sigma^{\rm y}_{\rm zx}  = \sigma^{\rm z}_{\rm xy} = -\sigma^{\rm x}_{\rm zy}
=-\sigma^{\rm y}_{\rm xz}  = -\sigma^{\rm z}_{\rm yx}$, i.e.\  cyclic permutation of the
indices gives no change while anticyclic permutation changes the sign.
In contrast to other derivations, there is obviously no need to artificially
introduce a spin-projected conductivity nor to make reference to the
conductivity tensor of a spin-polarized solid.  For a ferromagnetic metal with
fcc- or bcc-structure (\spacegroup{4/mm'm'}) with the magnetization along the
z-direction, the well-known shape of the conductivity tensor  $\matrg \sigma$
is obtained that reflects the anomalous Hall effect ($\sigma_{\rm xy}$) as well as
the magneto-resistance anisotropy ($\sigma_{\rm xx}\neq\sigma_{\rm zz}$) with the
symmetry relations $\sigma_{\rm xy} = - \sigma_{\rm yx}$ and $\sigma_{\rm xx} =
\sigma_{\rm yy}$.
As one notes the spin conductivity tensors show as for the non-magnetic case
off-diagonal elements that  represent the transverse spin conductivity.  This
implies  the occurrence of the spin Hall effect in ferromagnets that was
investigated recently for  diluted alloys\,\cite{ZCK+14a}.  However, more
elements show up as compared to the non-magnetic case since less symmetry
relations survive in the presence of a spontaneous magnetization.
Additionally, in contrast to the non-magnetic case also a longitudinal
spin-polarized conductivity ($\sigma^{\rm z}_{ii}$) occurs in a ferromagnet, that
for example gives rise to the spin-dependent Seebeck effect\,\cite{SBAW10}.  A
simple explanation for the corresponding longitudinal spin transport would be
based on Mott's two-current model assuming different conductivities for the
two spin channels.  However, it is well known that spin-orbit interaction leads
to a hybridization of the spin channels and influences even the longitudinal
conductivity of a ferromagnet this way\,\cite{BEV97}. Accordingly, it cannot be
ruled out that the longitudinal tensor elements $\sigma^{\rm z}_{ii}$ are not only
reflecting the spontaneous spin magnetization of the material but are to some
extent due to  spin-orbit coupling.

Indeed the scheme presented above leads for non-magnetic systems having low
symmetry not only to off-diagonal elements reflecting transverse spin
conductivity, i.e.\  the SHE, but also to diagonal elements reflecting
longitudinal spin transport, that was not observed so far.  For the two
magnetic Laue groups \spacegroup{4/m1'} and \spacegroup{2/m1'} for non-magnetic
solids considered in Table  \ref{tab-TENSOR}, a 4- and 2-fold, resp., rotation
axis is present. As a consequence longitudinal spin currents show up only with
spin polarization along this principal axis of rotation. 

To verify the results of our group theoretical approach independently we
calculated the full spin conductivity tensor for solids having different
structures corresponding to different magnetic Laue groups. This work employs a
computational scheme that has been used before for  numerical studies on the
SHE in non-magnetic transition metal alloys\,\cite{LGK+11}.  
Performing these calculations without making use of symmetry led numerically to 
a spin conductivity tensor that was always fully in line with the analytical 
group-theoretical results concerning the shape and degeneracies of the tensor.

\begin{figure}[thbt]
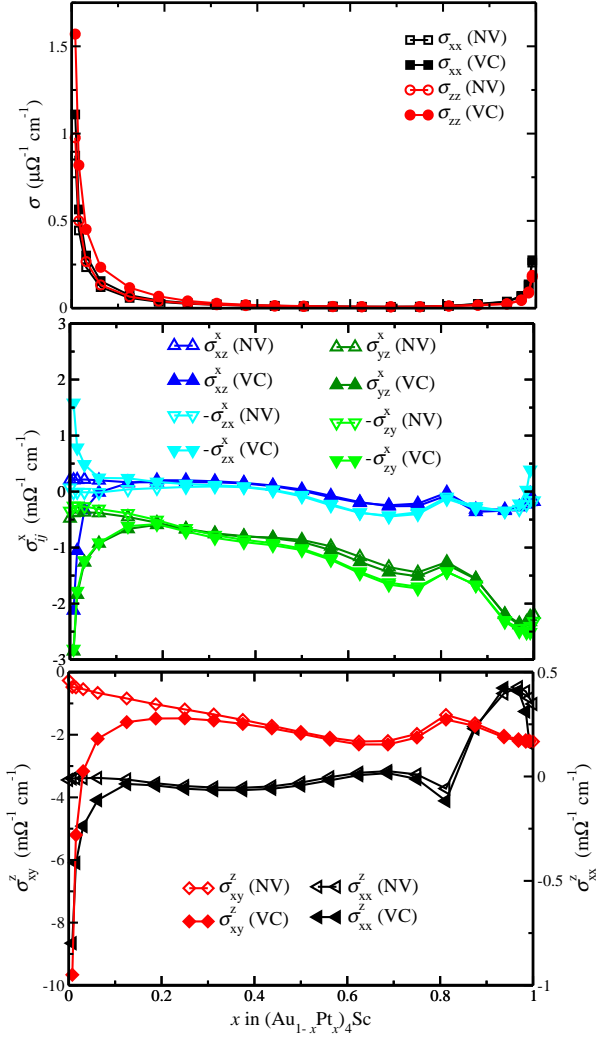

 \begin{center}
 \includegraphics[angle=0,width=0.8\linewidth,clip]{FIG1a_X4Sc_aa.eps}
 \vspace*{-0.05cm} 
 \includegraphics[angle=0,width=0.81\linewidth,clip]{FIG1b_X4Sc_xpol.eps}
 \vspace*{0.01cm} 
 \hspace*{6mm}\includegraphics[angle=0,width=0.91\linewidth,clip]{FIG1c_X4Sc_zpol.eps}
  \caption{\label{FIG-SIG}
Top: longitudinal conductivity  $ \sigma_{ii}   $ for
(Au$_{1-x}$Pt$_{\rm x}$)$_4$Sc as a function of the concentration $x$ calculated
without (NC) and with (VC) the vertex corrections.  Middle: transverse  spin
conductivities  $\sigma^{\rm  x}_{ij}$.  Bottom: transverse    and  longitudinal
spin conductivity $\sigma^{\rm z}_{\rm xy}$ and  $\sigma^{\rm z}_{\rm xx}$, respectively.
}
 \end{center}
\end{figure}

To get a first estimate of the order of magnitude of the longitudinal spin
polarized conductivity in non-magnets, calculations have been done for the
system (Au$_{1-x}$Pt$_{\rm x}$)$_4$Sc having the magnetic Laue group
\spacegroup{4/m1'} for varying Pt concentration $x$.  Fig.\ \ref{FIG-SIG} (top)
shows the corresponding electrical conductivity that is, in agreement with
Table \ref{tab-TENSOR}, diagonal and slightly anisotropic, i.e.\ $ \sigma_{\rm xx}
= \sigma_{\rm yy} \approx \sigma_{\rm zz}  $.
%
%
%
%
Furthermore,  the conductivities $ \sigma_{ii}   $ are strongly asymmetric
w.r.t.\ to the concentration $x$ when replacing Au with prominent
$sp$-character at the Fermi level by Pt with dominant $d$-character.
Furthermore, one notes a relatively strong impact of the vertex corrections on
the Au-rich side of the system ($ x\approx 0 $) while these are much less
important on the Au-poor side ($ x\approx 1 $).  This observation is well known
from binary transition metal alloys, like Cu$_{1-x}$Pt$_{\rm x}$\,\cite{BEWV94} or
Ag$_{1-x}$Pd$_{\rm x}$\,\cite{TSL+08}, where the dominance of $sp$-character
changes to $d$-character when $x$ is varied from 0 to 1.

The transverse spin conductivity $\sigma^{\rm x}_{ij}$ is shown in the middle panel
of Fig.\ \ref{FIG-SIG}     for  x-polarization of the spin.  As Table
\ref{tab-TENSOR} shows going from \spacegroup{m\bar 3m1'} to \spacegroup{4/m1'}
symmetry the relation  $\sigma^{\rm x}_{\rm yz} = - \sigma^{\rm x}_{\rm zy}$  disappears, 
i.e.\ the corresponding sub-tensor is not anti-symmetric anymore.  A symmetric
component, which is by definition not present in the ordinary SHE, indeed can
be seen in Fig.\ \ref{FIG-SIG} (middle) although the deviations are not very
pronounced.  In line one finds (except for $x \rightarrow 0$) for the
additional non-zero tensor elements   $\sigma^{\rm x}_{\rm xz} \approx -
\sigma^{\rm x}_{\rm zx}$. \tr{The first coefficient relates a spin current $j^x_x$ 
polarized in the direction of motion to an electric field $E_z$,} whereas 
$\sigma^{\rm x}_{\rm zx}$ describes a spin current $j^x_z$ transverse, but with
the spin polarization parallel to the driving electric field $E_x$. To our
knowledge the corresponding effects \tr{have} not been considered so far. \tr{Interestingly,} 
both elements occur simultaneously for a given magnetic Laue group or both are 
absent.  However, compared to the spin Hall-like elements $\sigma^{\rm x}_{\rm yz}$ 
and  $\sigma^{\rm x}_{\rm zy}$ they are smaller. For y-polarization of the spin 
the corresponding tensor elements are uniquely related to those for x-polarization
 according to Table \ref{tab-TENSOR} and for this reason not given here. The
 tensor elements $\sigma^{\rm z}_{ij}$ for z-polarization are given in the lower 
panel of Fig.\ \ref{FIG-SIG}.  In line with Table \ref{tab-TENSOR} they obey the 
symmetry relation  $\sigma^{\rm z}_{\rm xy} = - \sigma^{\rm z}_{\rm yx}$ (i.e. 
describing the pure SHE) and differ from  $\sigma^{\rm x}_{\rm yz} $.  This 
difference however is, except again for $x \rightarrow 0$, not very pronounced. 
In particular, $\sigma^{\rm x}_{\rm yz} $ and $\sigma^{\rm z}_{\rm xy} $ show a 
similar variation with concentration $x$ that differs clearly from that of the 
longitudinal spin conductivity $\sigma^{\rm z}_{\rm xx} $ shown as well in 
Fig.\,\ref{FIG-SIG} (bottom). Although this new type of tensor element is overall
 somewhat smaller in magnitude than the dominating transverse elements it has 
nevertheless the same oder of magnitude, especially in the Au-rich regime, and 
for that reason it should be possible to determine it experimentally.

\begin{figure}[thbt]
 \begin{center}
 \includegraphics[angle=0,width=0.8\linewidth,clip]{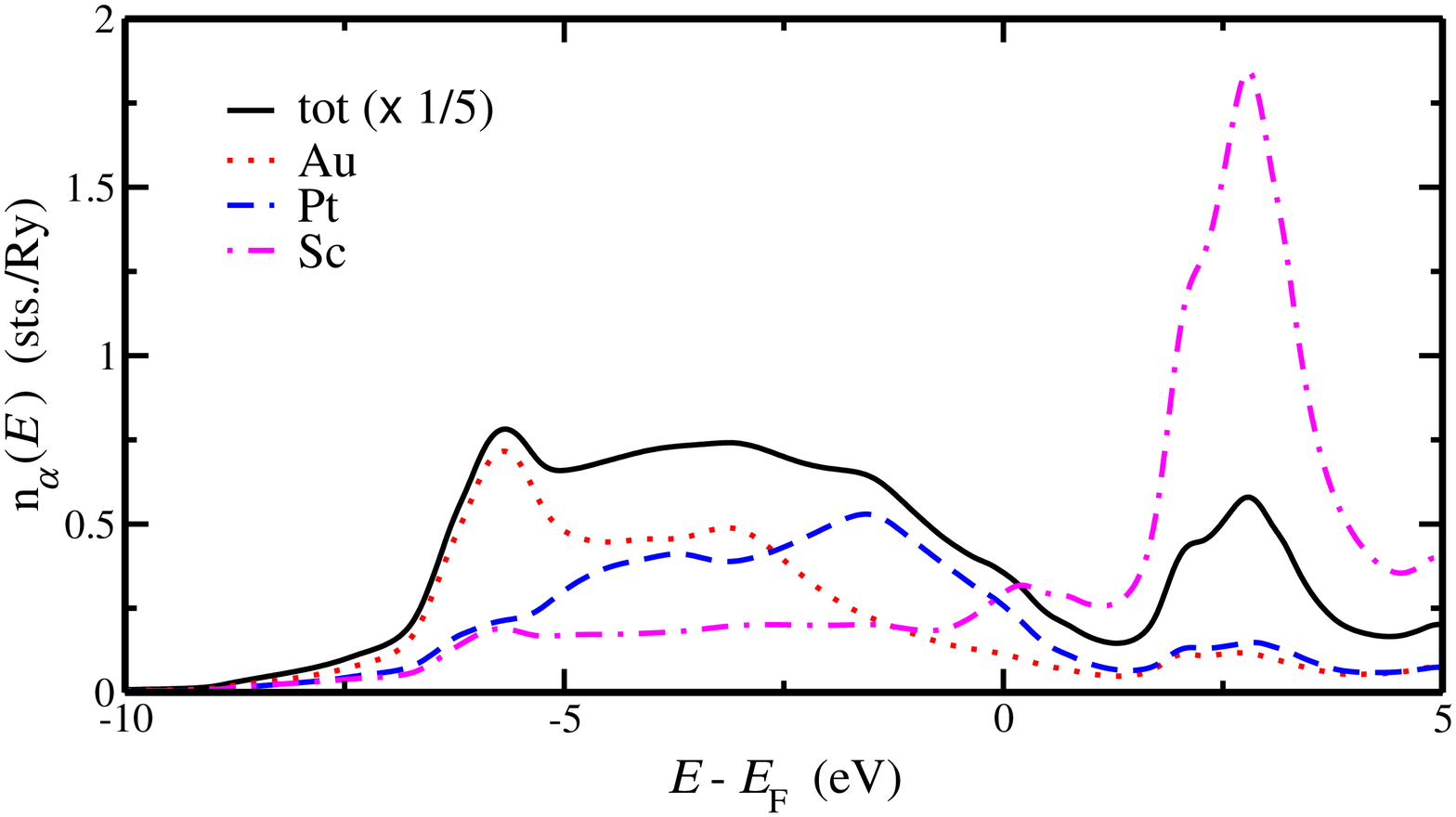}
 \includegraphics[angle=0,width=0.8\linewidth,clip]{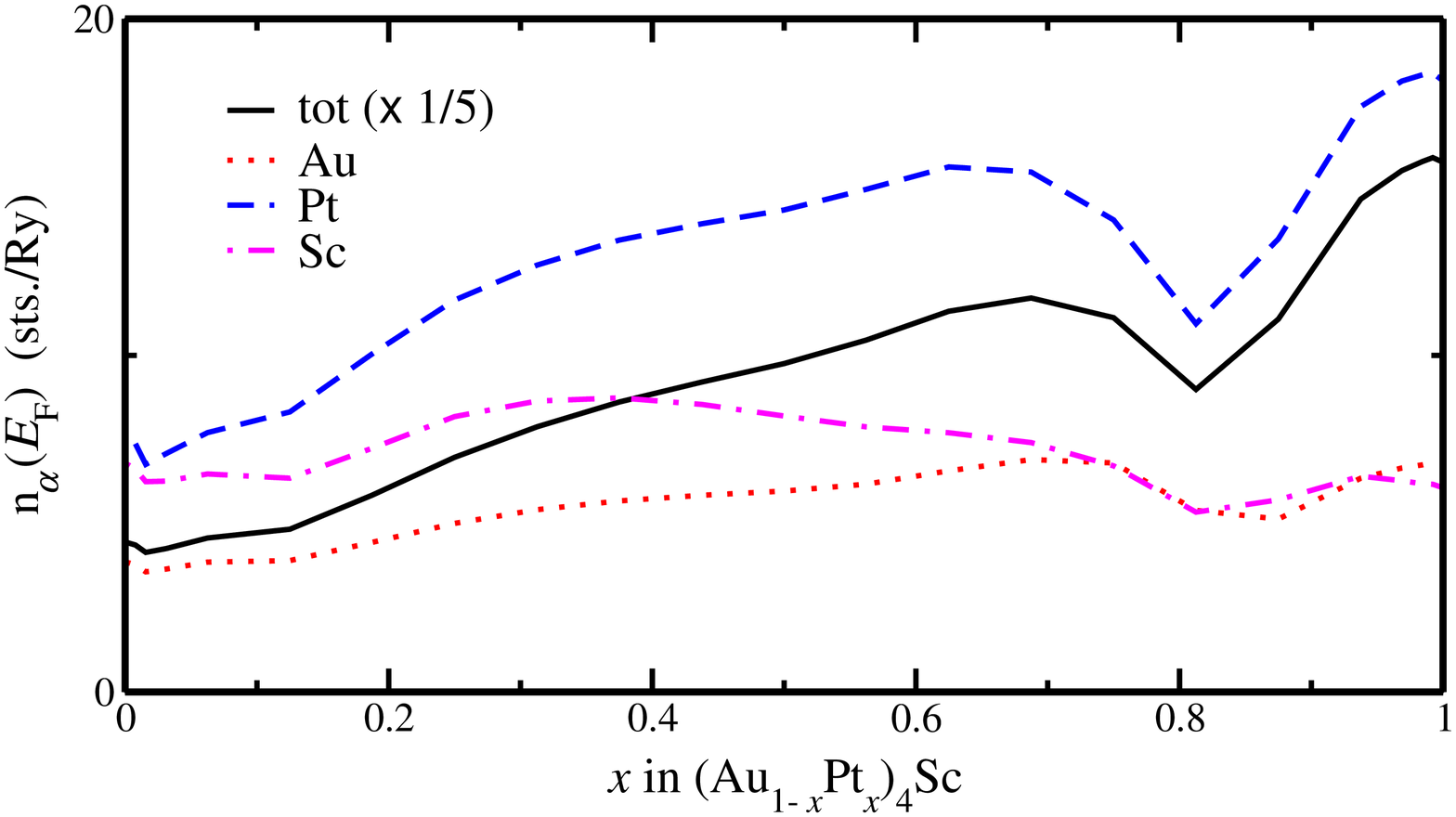}
  \caption{\label{FIG-DOS} 
Top: energy dependent component ($\alpha$) resolved DOS $n_\alpha(E)$  for
(Au$_{0.5}$Pt$_{0.5}$)$_4$Sc.  Bottom:  component resolved DOS $n_\alpha(E_F)$
at the Fermi energy $E_F$  for (Au$_{1-x}$Pt$_{\rm x}$)$_4$Sc as a function of the
concentration $x$.}
 \end{center}
\end{figure}

As can be seen in Fig.\ \ref{FIG-SIG}  the curves for the spin conductivity
tensor elements $\sigma_{ij}^k$ as function of the concentration $x$ are much
more structured than the electrical conductivity  $\sigma_{ii}$, i.e.\ they are
much stronger affected by the variation of the electronic structure with
composition. In particular the spin conductivities $\sigma_{ij}^k$ show
pronounced peaks or dips for $x \approx 0.8$.  This behavior can be related
to the variation of the density of states (DOS)  with $x$ as can be seen from
Fig.\ \ref{FIG-DOS}. It shows the component resolved DOS  $n_\alpha(E)$ as a
function of the energy $E$ for (Au$_{0.5}$Pt$_{0.5}$)$_4$Sc (top) and at the
Fermi energy $E_F$  for (Au$_{1-x}$Pt$_{\rm x}$)$_4$Sc as a function of the
concentration $x$ (bottom).
%
%
As mentioned above, at the Fermi energy the partial  DOS  $n_{\mbox{\tiny
Au}}(E_F)$ of Au  is  dominated by $sp$-states while that of Pt has dominant
$d$-character. The pronounced dip of the Pt DOS  $n_{\mbox{\tiny Pt}}(E_F)$ at
$x\approx 0.8$ is apparently responsible for the prominent features in the
spin conductivity curves   shown in Fig.\ \ref{FIG-SIG} (middle and bottom
panels).

As mentioned before, for the longitudinal conductivity  $\sigma_{ii}$ inclusion
of the vertex corrections has primarily an impact at the Au-rich side of the
system.  The same behavior is found for the transverse  ($\sigma^{k}_{ij} $)
as well as the longitudinal ($\sigma^{\rm z}_{ii} $) spin conductivity components.
For the transverse spin Hall conductivity it could be demonstrated that the
contribution connected with the vertex corrections corresponds to the so-called
extrinsic contribution that is primarily caused by the skew scattering
mechanism\,\cite{LGK+11,ZCK+14a}.  The very similar dependence of
$\sigma^{k}_{ij} $ and $\sigma^{\rm z}_{ii} $ on the vertex corrections suggests
that this applies also for the longitudinal spin conductivity.

\smallskip

In summary, a group-theoretical scheme has been presented that allows
determining the shape of response tensors relevant for the field of
spintronics. Application to the spin conductivity tensor gave a sound and
model-independent explanation for the occurrence of the transverse tensor
elements responsible for the  spin Hall effect \tr{and two additional, closely related effects}. 
In addition it was found that for low symmetry longitudinal
elements show up in addition even for non-magnetic solids that were not
considered before.  Independent numerical investigations confirmed these
results and demonstrated for (Au$_{1-x}$Pt$_{\rm x}$)$_4$Sc that the  longitudinal
spin conductivity may be in the same order of magnitude as the transverse one.
It should be noted that the discussion of the spin conductivity tensor
was referring to the $dc$-limit $\omega=0$. However, the tensor forms
given in Table~\ref{tab-TENSOR} also hold for finite frequencies, implying
the occurrence of the $ac$-counterparts to the discussed effects.
In addition, the formalism is applicable to numerous other linear response phenomena
as e.g. the AHE, anisotropic magneto-resistance (AMR), the Edelstein 
effect\,\cite{AL89a,Ede90}, Gilbert damping\,\cite{EMKK11}, spin-orbit torques\,\cite{GMA+13a}, 
etc.. Furthermore, using the fact that the operators for electrical and heat 
currents share the same transformation properties the presented formalism can be
 applied to spincaloritronic phenomena as well.

\smallskip

\begin{acknowledgements}
This    work    was   supported    financially    by   the    Deutsche
Forschungsgemeinschaft (DFG) priority program SPP\,1538 (Spin Caloric Transport)
and the SFB 689 (Spinph\"anomene in reduzierten Dimensionen). Discussions with
Ch.\,Back and H.\,H\"ubl are gratefully acknowledged.
\end{acknowledgements}


%

\end{document}